\documentclass[twocolumn,linenumbers]{aastex631}

\usepackage{threeparttable}

\begin{document}
\nolinenumbers
\title{A possible jet and corona configuration for Swift J1727.8--1613 during the hard state}

\author[0000-0002-5554-1088]{Jing-Qiang Peng\textsuperscript{*}}
\email{pengjq@ihep.ac.cn}
\affiliation{Key Laboratory of Particle Astrophysics, Institute of High Energy Physics, Chinese Academy of Sciences, 100049, Beijing, China}
\affiliation{University of Chinese Academy of Sciences, Chinese Academy of Sciences, 100049, Beijing, China}
\author{Shu Zhang\textsuperscript{*}}
\email{szhang@ihep.ac.cn}
\affiliation{Key Laboratory of Particle Astrophysics, Institute of High Energy Physics, Chinese Academy of Sciences, 100049, Beijing, China}

\author[0000-0001-5160-3344]{Qing-Cang Shui}
\affiliation{Key Laboratory of Particle Astrophysics, Institute of High Energy Physics, Chinese Academy of Sciences, 100049, Beijing, China}
\affiliation{University of Chinese Academy of Sciences, Chinese Academy of Sciences, 100049, Beijing, China}
\author[0000-0001-5586-1017]{Shuang-Nan Zhang}
\affiliation{Key Laboratory of Particle Astrophysics, Institute of High Energy Physics, Chinese Academy of Sciences, 100049, Beijing, China}
\affiliation{University of Chinese Academy of Sciences, Chinese Academy of Sciences, 100049, Beijing, China}

\author[0000-0001-8768-3294]{Yu-Peng Chen}
\affiliation{Key Laboratory of Particle Astrophysics, Institute of High Energy Physics, Chinese Academy of Sciences, 100049, Beijing, China}



\begin{abstract}
\nolinenumbers
Swift J1727.8--1613 is a black hole X-ray binary that differs from other black hole X-ray binaries in that it has an extra hard component in addition to a reflection component.
We perform spectral analysis with simultaneous Insight-HXMT, NICER and NuSTAR observations when the source was in the hard and hard intermediate states.
For the presentation of the extra components, we investigate the correlation between the {\tt relxill} parameters. We find that the correlation between the spectral index and the reflection fraction is consistent with MAXI J1820+070 when the jet dominates the reflection. This provides the second sample to have such a correlation during an outburst. 
Interestingly, when the reflection component is attributed to the corona, the spectral fit results in a small reflection fraction and the correlation between the spectral index and reflection fraction is in agreement with the overall trend built-in \cite{2023ApJ...945...65Y} with a large sample of outbursts from other X-ray binaries. 
Hence Swift J1727.8--1613 turns out to be the first sample to bridge the MAXI J1820+070 to the majority of X-ray binaries according to the dual correlations observed between the spectral index 
and the reflection fraction. We speculate that a configuration of a jet plus a hot inner flow may account for this peculiar outburst behavior of Swift J1727.8--1613.

\end{abstract}



\section{Introduction} \label{sec:intro}

A black hole low-mass X-ray binary (LMXRB) is composed of a companion star with a low mass ($\lesssim 1M_\odot$) and a black hole (BH). 
In these systems, BH accretes material from their companion star through the Roche lobe, leading to X-ray emission \citep{1973Shakura}.
The low-mass black hole X-ray binary is typically only easily observed in the X-ray band when it enters an outburst. During an outburst, the system exhibits increased X-ray emission, making it more readily detectable by X-ray telescopes.

BH LMXBs undergo outbursts characterized by different spectral states, including the low/hard state (LHS), high/soft state (HSS), and intermediate state (IMS). Each spectral state occupies a distinct position on the hardness-intensity diagram (HID) and exhibits unique time and spectral characteristics \citep{2005Belloni,2009Motta}.
In the LHS, the non-thermal component dominates, seed photons from the accretion disk are Compton scattered by the hot electron cloud in the corona/jet, leading to the generation of hard X-ray radiation.
The reflective component is often observed as a result of primary X-rays illuminating the reflective materials such as disks. The reflection component consists of fluorescent emission lines and a hump in the energy range of approximately 20--40 keV.
The common model used to fit the reflected component is the {\tt relxill} family, which contains both primary and reflected radiation in a self-consistent manner \citep{2016Dauser}. 

In the LHS, the spectrum typically includes a thermal component from the disk and a non-thermal component. However, in some black hole X-ray binaries, two non-thermal components may also be observed, with one of them displaying a high-energy tail \citep{2021Kong,2021You, 2024ApJLP}.
In the HSS, the thermal radiation from the disk is dominant and the spectrum is softer, with a higher spectral index around 2.1-3.7 \citep{1997Esin}, and there can also be a weak reflective component, which is usually considered to be the self-irradiation of the disk.
The IMS serves as an intermediary between the LHS and HSS and can be further divided into hard and soft intermediate states \citep{2005Belloni}.

Swift J1727.8--1613 is a black hole low-mass X-ray binary discovered by the SWIFT/BAT telescope on August 24, 2023 \citep{2023GCN.34537....1P}. 
Follow-up observations were conducted in X-ray, optical, and radio wavelengths \citep{2023Nakajima,2023O'Connor,2023Castro-Tirado,2023Miller}.
The distance and orbital period of Swift J1727.8--1613 are about $2.7 \pm 0.3$ kpc and 7.6 h respectively \citep{2024A&A...682L...1M}.
The system has a medium-high inclination and harbors a highly rotated BH \citep{2023Draghis,2024ApJLP,2024arXiv240501498C}.
\cite{2024ApJLP} analyzed the spectrum in the broad energy band with simultaneous NICER, NuSTAR, and Insight-HXMT observations and found an extra nonthermal component in addition to the reflection component.
INTEGRAL observations revealed that Swift J1727.8--1613 exhibits a high-energy tail at \textgreater 100 keV throughout the entire hard intermediate state (HIMS) \citep{2023ATel16238....1C,2024A&A...688L...5B}.
\cite{2024ApJ...971L...9W} discovered the largest resolved continuous jet in an X-ray binary ever observed through the Very Long Baseline Array (VLBA) and the Long Baseline Array (LBA) observations.

In this paper, we analyze nearly simultaneous observations of Swift J1727.8--1613 from Insight-HXMT, NICER, and NuSTAR. The correlation between the parameters of the reflective components is investigated, with emphasis on its evolution within outburst and the comparison with other XRBs. 
In Section \ref{obser}, we describe the observations and data reduction. The detailed results are presented in Section \ref{result}. The results are then discussed, and the conclusions are presented in Section \ref{dis}.

\section{Observations and Data reduction}
\label{obser}

\begin{table*}[htbp]
    \centering
		\caption{NICER, NuSTAR and Insight-HXMT    observations of Swift J1727.8--1613  during the 2023 outburst. }
		\begin{tabular}{cccccc}
		  \hline
		   \hline
		   NICER & Observed date & Exposure Time \\ObsID& (MJD) & (s) \\ \hline
      6703010101 &60185.55 & 30700 \\ 
      6703010102 &60188.58 & 748 \\
      6203980111 &60191.04 & 9693  \\
      6750010501 &60194.79 & 31090\\ 
     6750010502 &60195.03& 30910 \\ 
       \hline \hline
         NuSTAR &  Observed date & FPMA exposure time & FPMB exposure time
         \\ ObsID& (MJD) & (s) & (s) 
         \\ \hline
       90902330002 &  60185.42 &923& 1012 \\ 
       80902333002 &  60188.56 &1519&1667 \\ 
       80902333004 & 60191.10 &1256& 1369 \\ 
       80902333006 &  60194.78 &673.5& 733.2 \\  
      80902333008 &  60195.05 &543.2& 589.5 \\ 
       
       		  \hline \hline
           
         Insight-HXMT &  Observed date & ME exposure time & HE exposure time
         \\ ObsID& (MJD) & (s) & (s) 
       \\ \hline
      P061433800302 &60185.46& 2942 & 1917  \\ 
      P061433800411 &60188.50& 1912 &2130  \\
      P061433800601 &60191.06& 3842 &3884  \\
      P061433800807 &60194.85& 547.6 &193.9  \\
      P061433800901 &60195.09& 3332 &2298  \\
    \hline
        \label{observ-ninuhx} &     

    \end{tabular}

\end{table*}

\begin{table*}[]
    \centering
		\caption{Insight-HXMT and NICER observations of Swift J1727.8--1613 during the 2023 outburst. }
		\begin{tabular}{ccccccc}
		  \hline
              \hline
        Insight-HXMT &  Observed date & ME exposure time & HE exposure time&NICER& Observed date & Exposure Time 
         \\ ObsID& (MJD) & (s) & (s) &ObsID &(MJD)&(s)
       \\ \hline
       P061433800101 & 60181.34  &2548& 2305& 6203980101 &60181.03 & 612 \\ 
       P061433800110 & 60182.56 &1772 & 2824& 6203980102 &60182.59 & 2622  \\ 
       P061433800203 &60183.35 &2147 & 1696 &6203980103 &60183.36 & 1281 \\ 
       P061433800212 & 60184.54 &2222 & 2207&6203980104 &60184.65 & 1852  \\ 
       P061433800301 & 60185.30  &2610 & 2145& 6203980105 &0185.04 & 3627 \\ 
       P061433800405 & 60187.70 &1421 & 2103&6203980107 &60187.74 &1620  \\ 
       P061433800407 & 60187.97  &613.5 & 261& 6203980108 &60188.01 & 8232 \\ 
       P061433800501 &60189.08 &2470 & 3597&6203980109 &60189.04 & 12550 \\ 
       P061433800508 & 60190.02 &1832 & 1670&6750010201 &60190.01 & 7206  \\ 
       P061433800512 & 60190.55  &1618 & 2688& 6203980110 & 60190.52& 6659  \\ 
       P061433800603 &60191.34 &1922 & 1507&6703010103 &60191.35 & 2788 \\ 
       P061433800608 & 60192.00 &2099 & 2248& 6203980112 &60192.06 & 6770  \\ 
       P061433800616 & 60193.06 &2676 & 3938&6750010301 &60193.03 & 7351  \\ 
       P061433800618 & 60193.32  &1761 & 2058& 6203980112 & 60193.42& 5400 \\ 
       P061433800801 & 60194.04 &2625 & 3940& 6203980114 &60194.00 &7308  \\ 
       P061433800905 &60195.64 &612.5 & 1027&6203980115 &60195.61 & 171 \\ 
       
    \hline
        \label{observation-nicer-hxmt}    

    \end{tabular}

\end{table*}

\subsection{NICER}

The Neutron Star Interior Composition Explorer (NICER) is an International Space Station (ISS) payload, which was launched by the Space X Falcon 9 rocket on 3 June 2017 \citep{2016Gendreau}. 

NICER has conducted multiple observations of Swift J1727.8--1613, and simultaneous observations with Insight-HXMT and NuSTAR can extend the investigation of spectral properties to high-energy bands (see Table \ref{observ-ninuhx} and \ref{observation-nicer-hxmt}).
NICER data are reduced using the standard pipeline tool {\tt nicer}l2\footnote{\url{https://heasarc.gsfc.nasa.gov/lheasoft/ftools/headas/nicerl2.html}}. 
We extract light curves using {\tt nicer}l3-lc\footnote{\url{https://heasarc.gsfc.nasa.gov/docs/software/lheasoft/ftools/headas/nicerl3-lc.html}}.
To extract the spectrum, we utilize {\tt nicer}l3-spect\footnote{\url {https://heasarc.gsfc.nasa.gov/docs/software/lheasoft/help/nicerl3-spect.html}}, employing the "{\tt nibackgen3C50}\footnote{\url{https://heasarc.gsfc.nasa.gov/docs/nicer/analysis_threads/background/}}" model to estimate the background for spectral analysis.
 {\tt Nicerl3-spect} also applies the systematic error using niphasyserr automatically.
For the spectrum fitting, we choose an energy range of 1--10 keV, because NICER below 1 keV has significant residual due to calibration problems \citep{2020ApJ...899...44W}.

\subsection{NuSTAR}

The Nuclear Spectroscopic Telescope Array (NuSTAR) is the first mission to use focusing telescopes to image the sky in the high-energy X-ray (3--79 keV) region of the electromagnetic spectrum, which was launched at 9 am PDT, June 13, 2012 \citep{2013Harrison}.
The NuSTAR observations conducted simultaneously with NICER and Insight-HXMT (see Table \ref{observ-ninuhx}), are adopted to investigate the spectral characteristics of Swift J1727.8--1613 across a wide energy range.
We extract NuSTAR filtered data using the standard pipeline program {\tt nupipeline}, 
Since Swift J1727.8--1613  is very bright we make the parameter statusexpr="(STATUS==b0000xxx00xxxx000)\&\&(SHIELD\par==0)" followed by using nuproduct to extract the spectrum. The spectrum is extracted from a 120$''$ circle region centered on the source and the background is generated from a 60$''$ circle region away from the source. We adopt both FPMA and FPMB data for spectral analyses.

\subsection{Insight-HXMT}

Insight-HXMT is the first Chinese X-ray astronomy satellite and was successfully launched on 2017 June 15 \citep{2014Zhang, 2018Zhang, 2020Zhang}.
It carries three scientific payloads: the Low Energy X-ray Telescope (LE, SCD detector, 1--15 keV), the Medium Energy X-ray Telescope (ME, Si-PIN detector, 5--35 keV), and the High Energy X-ray Telescope (HE, phoswich NaI (CsI), 20--250 keV) \citep{2020Chen,2020Cao,2020Liu}. This enables observations in the energy range of 1--250 keV.

Insight-HXMT started to observe Swift J1727.8--1613  on August 25, 2023. 
We chose observations taken almost simultaneously with NICER and NuSTAR for a joint spectral analysis (see table \ref{observ-ninuhx} and \ref{observation-nicer-hxmt}).
We extract the data from ME and HE using the Insight-HXMT Data Analysis software {\tt{HXMTDAS v2.05}}. LE data are not included due to the temporary calibration issues. Since the LE is an SCD detector, there is a pile-up effect which is in the process of being calibrated.
The data are filtered with the criteria recommended by the Insight-HXMT Data Reduction Guide {\tt v2.06}\footnote{\url{http://hxmtweb.ihep.ac.cn/SoftDoc/847.jhtml}}.
The energy bands considered for spectral analysis are ME 10--28 keV and HE 28--120 keV. One percent systematic error is added to data, and errors are estimated via  Markov Chain Monte-Carlo (MCMC) chains with a length of 20000.

\section{Results}
\label{result}

\subsection{ Light curve and Hardness-Intensity Diagram}
\label{lchard}
\begin{figure*}
	\centering
	\includegraphics[angle=0,scale=0.8]{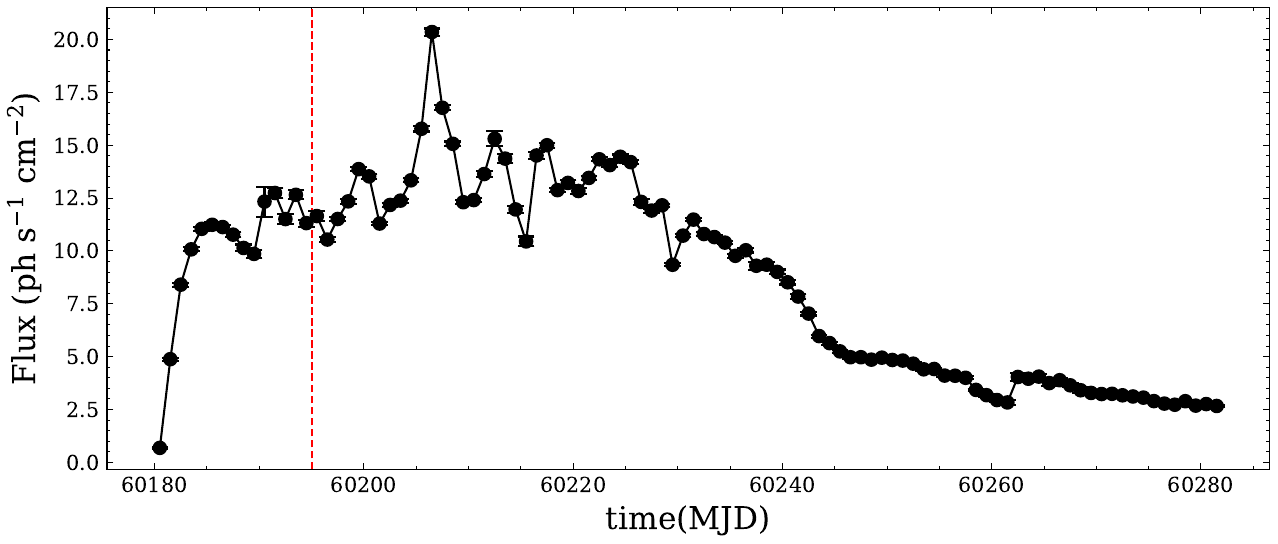}
	\caption{The light curves of Swift J1727.8--1613 observed by MAXI in the 2--20 keV energy range during the 2023 outburst.  The data we used is before the red vertical line. }
	\label{lcurve}
\end{figure*}

\begin{figure}
	\centering
	\includegraphics[angle=0,scale=0.6]{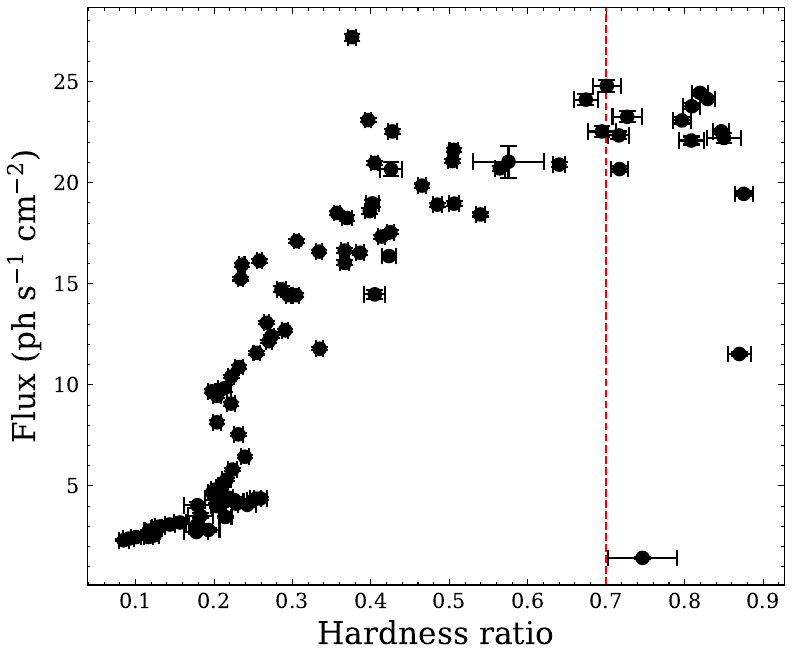}
	\caption{The MAXI hardness-intensity diagram of Swift J1727.8--1613 with time bin 1 d, where the hardness is defined as the ratio of 4--10 keV to 2--4 keV count rate.   The data we used is a hardness ratio greater than 0.7.}
	\label{hid}
\end{figure}

As shown in Figure \ref{lcurve}, the flux of Swift J1727.8--1613 exhibits a rapid increase.
Swift J1727.8--1613 was observed almost simultaneously by NICER, NuSTAR, and Insight-HXMT several times (see Table \ref{observ-ninuhx} and \ref{observation-nicer-hxmt}).
To investigate the timing of these observations during the Swift J1727.8--1613 outburst, we calculate the hardness ratio, defined as the ratio of 4--10 keV to 2--4 keV count rate. 
These simultaneous observations are located in hard and hard intermediate states on the HID and have a hardness ratio greater than 0.7 (Figure \ref{hid}).

\subsection{The spectral analysis}
\label{spec}

As shown in Table \ref{observ-ninuhx},  there are five sets of simultaneous joint observations from Insight-HXMT, NICER, and NuSTAR. Accordingly, joint spectral fittings are carried out for each observational set. 
The model for which we carry out the spectrum fitting is consistent with model 1 of \cite{2024ApJLP} as {\tt plabs*tbabs(diskbb+relxill+cutoffpl)}.

We utilize the {\tt plabs} to account for the variance in spectral index between NICER, NuSTAR, and Insight-HXMT \citep{2021Zdziarski}. We have set the values of K and $\Delta \Gamma$  to 1 and 0 for NICER, while for Insight-HXMT and NuSTAR, these parameters are free during the fitting.
The interstellar absorption and multitemperature blackbody components we fit with {\tt tbabs} and {\tt diskbb}, respectively \citep{1984Mitsuda,2000Wilms}.
{\tt Cutoffpl} was used to fit non-thermal components other than {\tt relxill}
We add the relativistic reflection model RELXILL to fit the reflection component in the spectrum \citep{2016Dauser}.
{\tt Relxill} is the standard relativistic reflection model, and the incident spectrum is the standard high-energy cutoff power-law, which contains both primary and reflected radiation in a self-consistent manner.

In the fitting, we fix the spin and inclination of the {\tt relxill} at 0.99 and 40° respectively. 
We fix the emission index of the disk, the truncation radius $R_{\rm br}$, and the outer disk radius Rout at 3, 100, and 400.  And other {\tt relxill} parameters are free parameters in the spectral fits.
Based on \cite{2024ApJLP} results, the model exhibits coupling, resulting in two sets of fitting results and parameters. 
Swift J1727.8--1613 shows an additional component in the spectrum apart from the reflection one.
This additional component can be described as a medium-energy hump or a high-energy tail.
For each of these two conditions, we investigate the correlation between the parameters of the reflection.
We use {\tt cflux} to calculate fluxes in the 1--120 keV energy band for {\tt diskbb, relxill and cutoffpl}.

As shown in Table \ref{observation-nicer-hxmt}, there are 16 nearly simultaneous observations of Swift J1727.8--1613 by NICER and Insight-HXMT.
We adopt the above model and the same parameter settings for the joint fitting of the spectra of NICER and Insight-HXMT. The two sets of fitting results and parameters are shown in Tables \ref{T-R-H}, \ref{T-R-T}.

\subsubsection{Reflective fitting of medium energy hump}
\label{hump}

\begin{figure*}
	\centering
	\includegraphics[angle=0,scale=0.7]{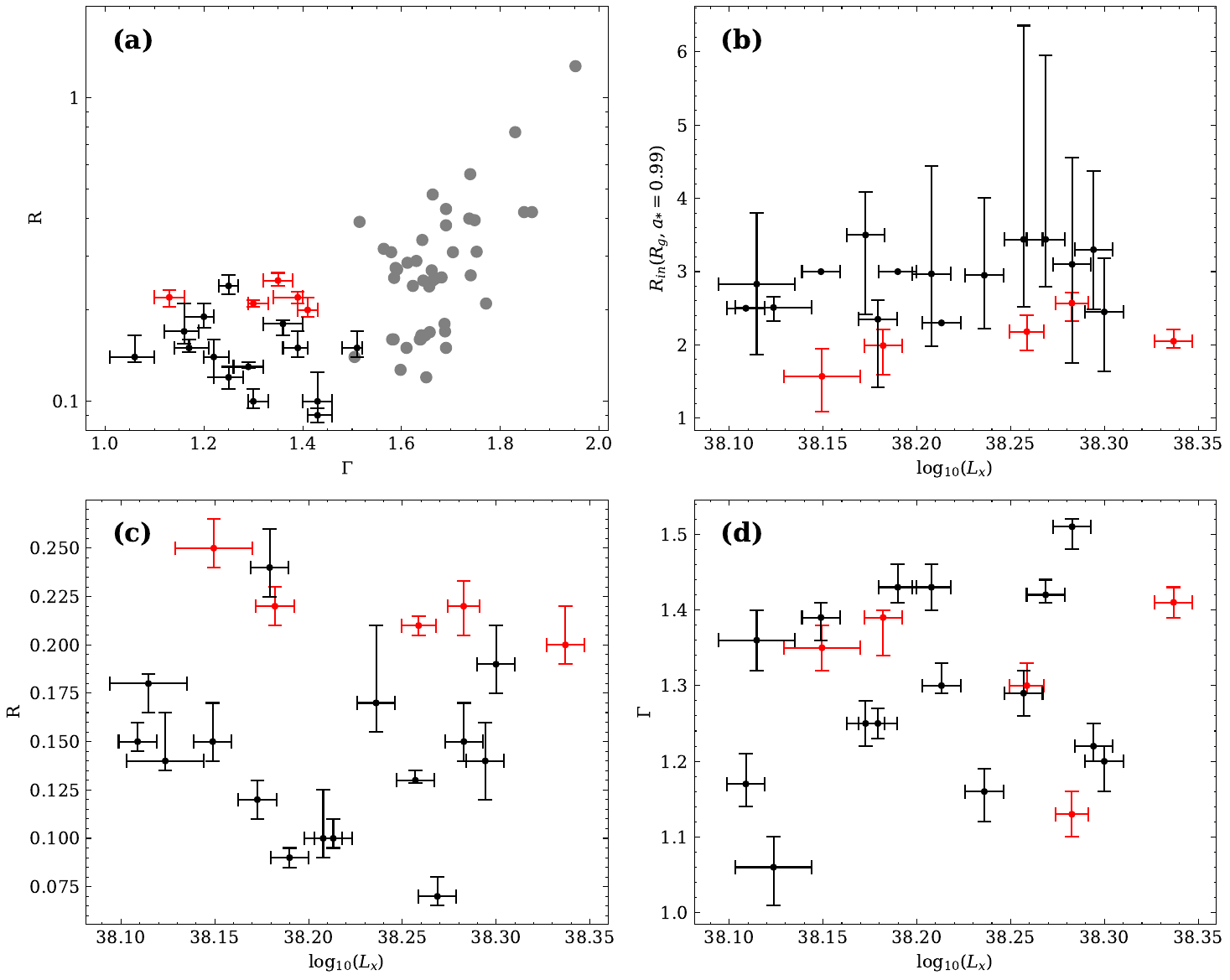}
	\caption{The relationship between {\tt relxill} parameters.
 The black dots indicate the joint fit results from NICER and Insight-HXMT. The red dots correspond to the joint fit results of NICER, NuSTAR, and Insight-HXMT.
 (a) The relationship between $\Gamma$ and $R$. The gray dots represent the data from Figure 2 in \cite{2023ApJ...945...65Y} for a large sample of XRB outbursts.
 (b) (c) (d) The relationships between $R_{in}$, $R$, $\Gamma$ and luminosity $L_{x}$ of the {\tt relxill} component in 1--120 keV for Swift J1727.8--1613.}
	\label{hump-relation}
\end{figure*}

When the extra components in addition to the reflection component exhibit a high-energy tail, i.e., the reflection is used to fit the medium-energy hump,
the reflection fraction $R$ and spectral index $\Gamma$ relationship is shown in (a) of Figure \ref{hump-relation}.
For Swift J1727.8--1613, the correlation between $R$ and $\Gamma$ 
is weak, with a correlation coefficient of -0.20 and a significance of only 0.86 $\sigma$ due to the limited range of variation in the spectral index. 
However, when expanding the sample of black holes and incorporating them into the relationship between $R$ and $\Gamma$ for other black hole X-ray binaries which was reported in \cite{2023ApJ...945...65Y},  
the data were sampled 100,000 times within the margin of error and the average correlation coefficient and significance improve to 0.50 and 4.31 $\sigma$, respectively, rejecting the null hypothesis that $R$ and $\Gamma$  are not linearly correlated. This indicates that there is a significant positive correlation between $R$ and $\Gamma$  from the enlarged XRB sample. 
As shown in (b) (c) (d) of figure \ref{hump-relation},
the relationships between $R_{in}$, $R$, $\Gamma$ and the luminosity $L_{x}$ of the {\tt relxill} component in 1--120 keV for Swift j1727.8--1613, do not exhibit a strong correlation. This is attributed to the limited range of luminosity variations.

\subsubsection{Reflective fitting of high-energy tail}
\label{tail}

\begin{figure*}
	\centering
	\includegraphics[angle=0,scale=1]{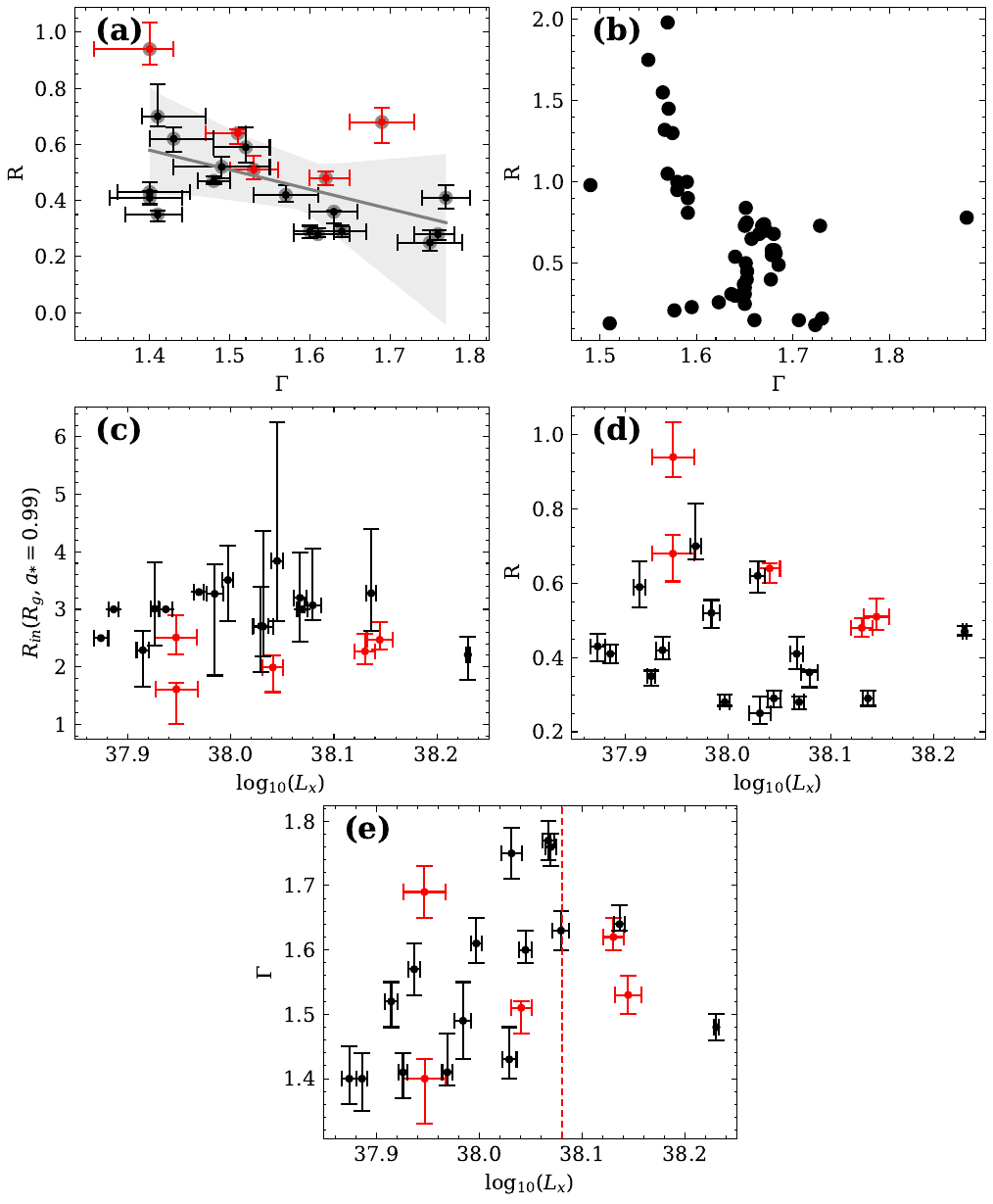}
	\caption{The relationship between {\tt relxill} parameters.
 The black dots indicate the joint fit results from NICER and Insight-HXMT. The red dots correspond to the joint fit results of NICER, NuSTAR, and Insight-HXMT.
 (a) The relationship between $\Gamma$ and $R$. The grey lines are the fits with linear slopes of $-0.70 \pm 0.28$.  Shaded areas are 95\% confidence intervals. (b) The dots represent the data from MAXI J1820+070 in \cite{2023ApJ...945...65Y}. 
 (c) (d) (e) The relationships between $R_{in}$, $R$, $\Gamma$ and luminosity $L_{x}$ of the {\tt relxill} component in 1--120 keV for Swift J1727.8--1613.}
	\label{rela-tail}
\end{figure*}

When the extra components in addition to the reflection component exhibit a medium-energy hump, i.e., the reflection is used to fit the high-energy tail,
the reflection fraction $R$ and spectral index $\Gamma$ relationship is shown in (a) of Figure \ref{rela-tail}.
We apply the linear regression function to fit the reflection fraction 
 $R$ and spectral index $\Gamma$, resulting in a slope of  $-0.70 \pm 0.28$. This anti-correlation is consistent with the MAXI J1820+070 (see Figure 3 in the \cite{2023ApJ...945...65Y}).
The resulting reflection fraction is much higher than that derived from fitting a medium-energy hump with {\tt rexill}.

 As shown in (c) (d) (e) of figure \ref{rela-tail},
the relationships between $R_{in}$, $R$ and the luminosity $L_{x}$ of the {\tt relxill} component in 1--120 keV for Swift J1727.8--1613, do not exhibit strong correlations. This is attributed to the limited range of luminosity variations. 
However, as the luminosity increases, there is a transition from a positive correlation to an inverse correlation between $\Gamma$ and $L_{x}$ at about 1.2$\times10^{38}~erg~s^{-1}$. 
This evolutionary trend aligns with that observed in MAXI J1820+070 (see Figure 8 in the \cite{2023ApJ...945...65Y}).
However, we fit this with linear and segmented linear and calculate the significance using {\tt ftest}, and we find the significance to be only 2.72 $\sigma$.

\section{discussion and conclusion}
\label{dis}

We have performed a joint spectral analysis of Swift J1727.8--1613 with nearly simultaneous observations from NICER, NuSTAR and Insight-HXMT.
Swift J1727.8--1613 has two non-thermal components.
We utilize {\tt relxil} to individually fit the two components and then investigate the correlation between the parameters of {\tt relxil} and find that the parameters exhibited two different correlations when {\tt relxill} fitted to different components may be due to the better data quality of Swift J1727.8--1613, and coexistence of the strong jet emission, which make Swift J1727.8--1613 distinctive from other XRBs. 
We compare the correlation of the two sets of parameters obtained from the swift J1727.8--1613 spectral fit with other XRBs to establish a possible configuration of jets and corona/hot inner flows, i.e., jets contributing the main reflections and corona/hot inner flows contributing the weak reflections.

In the hard state, the spectrum of a black hole X-ray binary typically consists of two components: the multi-temperature blackbody emission from the disk and non-thermal emission produced by the seed photons of the disk after Compton scattering by the hot electrons cloud of the corona. 
The joint spectral analysis suggests two possibilities for the presentation of the extra spectral component of Swift J1727.8--1613.
One possibility is the observation of a hard excess at high energies in the spectrum of Swift J1727.8--1613, in addition to the reflection component.
Another possibility is that the extra spectral component could be a medium-energy hump. We investigate the correlation between the parameters of the {\tt relxill} model under various conditions.

When the $E_{\rm cut}$ of the {\tt relxill} is relatively high, the reflection component comes mainly from the jet above the disk. The jet has higher speed and energy relative to the coronal electrons, and it is easier for the photons to gain higher energy after Comptonization \citep{2010LNP...794.....B}. The relationship between the parameters of {\tt relxill} is shown in Figure \ref{rela-tail}.
Panel (a) of Figure \ref{rela-tail} shows the relationship between the spectral index and the reflection fraction. A clear inverse correlation between the reflection fraction and the spectral index is observed, with a slope of $-0.70 \pm 0.28$. This finding is consistent with the results obtained for MAXI J1820+070 ((b) of Figure \ref{rela-tail}) and in contrast to the other BH XRBs ((a) of Figure \ref{hump-relation}). This provides so far the second sample to manifest such an anti-correlation.  Swift J1727.8--1613 shares features with MAXI J1820+070 in aspects of e.g. hard excess, strong radio emission, etc. 
For MAXI J1820+070, a powerful jet exists in the hard state.
From the spectral analysis, \cite{2021You} employed two reflection components to fit the spectrum of MAXI J1820+070 with Insight-HXMT data. Both illuminators were believed to originate from the jet.   
For Swift J1727.8--1613,  \cite{2024ApJLP} performed a joint spectrum analysis with NICER, Insight-HXMT and NuSTAR observations and saw a hard excess at the high energy band.
INTEGRAL observations revealed that Swift J1727.8--1613 exhibits a high-energy tail at \textgreater 100 keV throughout the entire hard intermediate state (HIMS) \citep{2023ATel16238....1C,2024A&A...688L...5B}.

Jets are observed as radiation in the radio band because the radiation mechanism is primarily the spiral motion of charged particles in a magnetic field, which produces radiation. This synchrotron radiation predominantly occurs in the radio band.
So the flux of radio radiation is related to the strength of the jet.   We investigate the flux of radio emission from the BH XRBs in \cite{2023ApJ...945...65Y}, and we find that except for MAXI J1820+070, the radio emissions from the other BH XRBs are relatively as weak as about less than 10 mJy \citep{2001MNRAS.323..517B,2012MNRAS.421..468M,2018ATel11539....1T,2019MNRAS.485.5235A,2020ApJ...891...31X,2022MNRAS.512.5037D,2021PASA...38...45C,2022ApJ...932...38S,2024ApJ...969L..33Y}.
For Swift J1727.8--1613 the radio flux can exceed 100 mJy at 3, 5, 8 GHz. And \cite{2024ApJ...971L...9W} discovered the largest resolved continuous Jet in an X-ray binary ever observed through the Very Long Baseline Array (VLBA) and the Long Baseline Array (LBA) observations.
Therefore, we speculate that the unique anti-correlation between the spectral index and reflection fraction observed in MAXI J1820+070 and Swift J1727.8--1613 may suggest that for both sources the hard excess and as well the reflection component is dominated by the jet emission. 

We notice that the reflection component can as well be attributed to the corona hump. In this case, the reflection fraction can be rather small. Interestingly, by combining with other XRB outbursts as presented in \cite{2023ApJ...945...65Y},  spectral index and reflection fraction tend to have an overall positive correlation.  This leads Swift J1727.8--1613 to be the first sample to have dual faces in manifesting the reflection component during the outburst. Since there is no way to distinguish the two non-thermal components with a simple spectrum fit. 
Both the hard components of a hard excess and corona hump can account for the reflection components, but with a large different reflection fraction. Both can be consistent with evolution trends built with different XRB samples \citep{2023ApJ...945...65Y}. 
To investigate this further, we simulate NuSTAR spectra with strong and weak jets using the {\tt fakeit} method and perform fitting analyses.
We first simulate the input of a strong jet with its contributing reflection, and then we fit this spectrum with a corona, and we find that there is a weak reflection in the residual, so we add a {\tt relxill} to fit the weak reflection in the residual, and we find that the spectrum can be fitted very well and that the norm and reflection fraction of this component is low relative to the input reflection, which is consistent with the use of corona to fit the reflections in Swift J1727.8--1613.
This is most likely because the Compton hump induced by the strong jet can partially merge into the corona component once fitting the reflection with the corona instead of the jet. 
We also simulate a weak jet-contributing reflection and add a coronal component that does not contribute to the reflection. 
In this case, by fitting the reflection with the corona component instead, we do find that 
the reflection fraction of the {\tt relxill} can be relatively large which is consistent with the results of other black hole XRBs in \cite{2023ApJ...945...65Y}.

To account for the distinctive behavior showing up in Swfit J1727.8--1613, we propose a potential configuration of the corona/jet for a black hole XRB in the hard state.
In a black hole XRB, during the hard state, a jet typically exists over a hot inner flow (corona). In the case of sources with strong jets, such as Swift J1727.8--1613 and MAXI J1820+070, the reflections are predominantly from the jets. The hot inner flow/corona, which has a relatively small solid angle with respect to the disk, contributes less to the reflections.
The dual faces of Swift J1727.8--1613 may bridge between  MAXI J1820+070 and other XRBs: it may be the case that the former is jet-dominated while the latter is corona-dominated with less jet contribution. Both jet and corona can present in the hard state of XRB outburst and, in case of having corona as a flat hot inner flow, reflection may mainly be contributed by the jet during the hard state. It is interesting that during the flare state of MJD 60197--60220, \cite{Cao} found that the source entered the very high state, with a large increase of the power law component and relatively weak reflection. There they argue that, once the source goes beyond the hard state, the jet is speeded up, with fewer photons reaching the disk. Therefore, the weak reflection attributed to the hot inner flow starts to show up. So the proposed jet/hot inner flow configuration is in line with \cite{Cao}  in a jet-evolution manner: a largely accelerated jet contributes less reflection, as was already suggested by \cite{2021You} during the outburst of MAXI J1820+070.

\begin{acknowledgments}
This work is supported by the National Key R\&D Program of China (2021YFA0718500), the National Natural Science Foundation of China under grants 12333007, 12027803, 12273030.
This work made use of data and software from the \textit{Insight}-HXMT mission, a project funded by the China National Space Administration (CNSA) and the Chinese Academy of Sciences(CAS). This work was partially supported by the International Partnership Program of the Chinese Academy of Sciences (Grant No.113111KYSB20190020).
This research has used software provided by data obtained from the High Energy Astrophysics Science Archive Research Center (HEASARC), provided by NASA’s Goddard Space Flight Center.
\end{acknowledgments}

%




\begin{table*}[]
    \centering
		\caption{The results of the joint fit, corresponding to Figure \ref{hump-relation}. }
  \resizebox{\textwidth}{0.5\textwidth}{
		\begin{tabular}{ccccccccccccccccccc}
		  \hline
		   \hline
          Insight-HXMT & MJD&$N_{\rm H}$&$T_{\rm in}$&$N_{\rm disk}$ &PhoIndex & HighEcut&$N_{\rm cutoffpl}$&$R_{\rm in}$&$\Gamma$&logxi&Afe&Ecut&feflfrac&$N_{\rm relxill}$&$F_{\rm disk}$&$F_{\rm cutoffpl}$&$F_{\rm relxill}$
\\NICER& &&& && &&&&&&&&&&&
     \\NuSTAR& &&& && &&&&&&&&&&&     
         \\ObsID &&$(10^{22} \rm cm^{-2})$&(keV)&($10^{4})$&&(keV)&&($R_{\rm g}$)&&&&(keV)&&&($10^{-8}$~erg~s$^{-1}$~cm$^{-2}$)&$(10^{-8}$~erg~s$^{-1}$~cm$^{-2}$)&$(10^{-8}$~erg~s$^{-1}$~cm$^{-2}$)
       \\ \hline

P061433800302&60185.46 &  $0.15_{-0.007}^{+0.008}$ & $0.46_{-0.008}^{+0.007}$ & $5.92_{-0.34}^{+0.29}$ & $1.62_{-0.018}^{+0.017}$ & $51.88_{-0.62}^{+0.31}$ & $12.52_{-0.9}^{+0.94}$ & $2.57_{-0.14}^{+0.24}$ & $1.13_{-0.03}^{+0.03}$ & $3.22_{-0.08}^{+0.07}$ & $1.0_{-0.18}^{+0.29}$ & $13.32_{-0.26}^{+0.28}$ & $0.22_{-0.03}^{+0.026}$ & $0.26_{-0.01}^{+0.01}$ & $0.137_{-0.004}^{+0.004}$ & $1.337_{-0.012}^{+0.012}$ & $1.917_{-0.036}^{+0.04}$ & \\
6703010101&60185.55 &   & &  & &  &  & & & &  & &  & &  &  &  & \\
90902330002&60185.42 &   & &  & &  &  & & & &  & &  & &  &  &  & \\
\hline

P0614330800411&60188.50 &  $0.20_{-0.004}^{+0.01}$ & $0.48_{-0.01}^{+0.005}$ & $6.03_{-0.27}^{+0.90}$ & $1.72_{-0.02}^{+0.02}$ & $62.09_{-2.71}^{+2.96}$ & $12.9_{-0.55}^{+1.13}$ & $2.18_{-0.22}^{+0.26}$ & $1.3_{-0.01}^{+0.03}$ & $3.25_{-0.05}^{+0.06}$ & $1.31_{-0.36}^{+0.39}$ & $14.76_{-0.22}^{+0.45}$ & $0.21_{-0.01}^{+0.01}$ & $0.26_{-0.005}^{+0.01}$ & $0.167_{-0.004}^{+0.004}$ & $1.167_{-0.011}^{+0.011}$ & $1.814_{-0.038}^{+0.038}$ & \\
6703010102&60185.58 &   & &  & &  &  & & & &  & &  & &  &  &  & \\
80902330002&60185.56 &   & &  & &  &  & & & &  & &  & &  &  &  & \\
\hline

P0614330800601&60191.06 &  $0.17_{-0.01}^{+0.01}$ & $0.48_{-0.01}^{+0.01}$ & $5.07_{-0.19}^{+0.80}$ & $1.5_{-0.04}^{+0.03}$ & $74.39_{-4.57}^{+6.94}$ & $4.0_{-0.38}^{+0.49}$ & $2.05_{-0.16}^{+0.09}$ & $1.41_{-0.02}^{+0.02}$ & $3.28_{-0.1}^{+0.04}$ & $0.88_{-0.2}^{+0.15}$ & $17.93_{-0.75}^{+0.78}$ & $0.2_{-0.02}^{+0.04}$ & $0.32_{-0.01}^{+0.01}$ & $0.147_{-0.003}^{+0.003}$ & $0.68_{-0.008}^{+0.008}$ & $2.172_{-0.051}^{+0.051}$ & \\
6203980111&60191.04 &   & &  & &  &  & & & &  & &  & &  &  &  & \\
80902330004&60191.10 &   & &  & &  &  & & & &  & &  & &  &  &  & \\
\hline

P0614330800807&60194.85 &  $0.20_{-0.01}^{+0.01}$ & $0.5_{-0.01}^{+0.01}$ & $4.06_{-0.25}^{+0.21}$ & $1.83_{-0.02}^{+0.04}$ & $81.46_{-3.72}^{+16.6}$ & $15.08_{-1.3}^{+0.89}$ & $1.57_{-0.38}^{+0.48}$ & $1.35_{-0.03}^{+0.03}$ & $3.27_{-0.05}^{+0.08}$ & $1.63_{-0.64}^{+0.05}$ & $15.93_{-0.86}^{+0.25}$ & $0.25_{-0.02}^{+0.03}$ & $0.2_{-0.01}^{+0.01}$ & $0.142_{-0.003}^{+0.003}$ & $1.131_{-0.01}^{+0.01}$ & $1.411_{-0.066}^{+0.066}$ & \\
6750010501&60194.79 &   & &  & &  &  & & & &  & &  & &  &  &  & \\
80902330006&60194.78 &   & &  & &  &  & & & &  & &  & &  &  &  & \\
\hline

P0614330800901&60195.09 &  $0.19_{-0.01}^{+0.01}$ & $0.48_{-0.01}^{+0.01}$ & $4.67_{-0.59}^{+0.37}$ & $1.81_{-0.03}^{+0.03}$ & $80.93_{-5.3}^{+8.77}$ & $12.81_{-0.98}^{+1.0}$ & $1.99_{-0.22}^{+0.4}$ & $1.39_{-0.05}^{+0.01}$ & $3.3_{-0.08}^{+0.16}$ & $1.0_{-0.06}^{+0.88}$ & $17.47_{-1.11}^{+0.42}$ & $0.22_{-0.02}^{+0.02}$ & $0.22_{-0.01}^{+0.01}$ & $0.132_{-0.003}^{+0.003}$ & $1.007_{-0.009}^{+0.009}$ & $1.521_{-0.035}^{+0.035}$ & \\
6750010502&60195.03 &   & &  & &  &  & & & &  & &  & &  &  &  & \\
80902330008&60195.05 &   & &  & &  &  & & & &  & &  & &  &  &  & \\
\hline

P061433800110&60182.56 &  $0.18_{-0.01}^{+0.01}$ & $0.36_{-0.01}^{+0.01}$ & $10.14_{-1.37}^{+1.65}$ & $1.62_{-0.04}^{+0.03}$ & $67.85_{-7.05}^{+3.2}$ & $7.18_{-0.65}^{+0.96}$ & $2.95_{-0.73}^{+1.06}$ & $1.16_{-0.04}^{+0.03}$ & $3.02_{-0.12}^{+0.24}$ & $1.37_{-0.3}^{+0.57}$ & $16.72_{-0.6}^{+0.59}$ & $0.17_{-0.015}^{+0.04}$ & $0.25_{-0.01}^{+0.01}$ & $0.056_{-0.003}^{+0.003}$ & $0.865_{-0.01}^{+0.01}$ & $1.722_{-0.04}^{+0.04}$ & \\
6203980102&60182.59 &   & &  & &  &  & & & &  & &  & &  &  &  & \\
& &   & &  & &  &  & & & &  & &  & &  &  &  & \\
\hline

P061433800203&60183.35 &  $0.21_{-0.01}^{+0.01}$ & $0.34_{-0.004}^{+0.01}$ & $16.03_{-3.33}^{+0.55}$ & $1.63_{-0.02}^{+0.03}$ & $61.8_{-3.17}^{+3.56}$ & $6.83_{-0.54}^{+0.37}$ & $2.35_{-0.93}^{+0.26}$ & $1.25_{-0.02}^{+0.02}$ & $2.81_{-0.01}^{+0.15}$ & $1.35_{-0.48}^{+0.67}$ & $16.2_{-0.32}^{+0.5}$ & $0.24_{-0.015}^{+0.02}$ & $0.22_{-0.01}^{+0.01}$ & $0.065_{-0.003}^{+0.003}$ & $0.765_{-0.007}^{+0.007}$ & $1.511_{-0.035}^{+0.035}$ & \\
6203980103&60183.36 &   & &  & &  &  & & & &  & &  & &  &  &  & \\
& &   & &  & &  &  & & & &  & &  & &  &  &  & \\
\hline

P061433800212&60184.54 &  $0.20_{-0.01}^{+0.01}$ & $0.41_{-0.01}^{+0.01}$ & $10.01_{-1.27}^{+1.01}$ & $1.66_{-0.03}^{+0.04}$ & $55.88_{-4.07}^{+3.68}$ & $12.41_{-1.5}^{+1.63}$ & $2.45_{-0.81}^{+0.73}$ & $1.2_{-0.04}^{+0.02}$ & $2.89_{-0.25}^{+0.19}$ & $2.92_{-0.72}^{+2.4}$ & $13.68_{-0.49}^{+0.29}$ & $0.19_{-0.015}^{+0.02}$ & $0.3_{-0.01}^{+0.02}$ & $0.127_{-0.003}^{+0.003}$ & $1.248_{-0.012}^{+0.011}$ & $1.995_{-0.046}^{+0.046}$ & \\
6203980104&60184.65 &   & &  & &  &  & & & &  & &  & &  &  &  & \\
& &   & &  & &  &  & & & &  & &  & &  &  &  & \\
\hline

P061433800301&60185.30 &  $0.19_{-0.01}^{+0.01}$ & $0.42_{-0.01}^{+0.01}$ & $9.59_{-1.51}^{+0.64}$ & $1.65_{-0.02}^{+0.03}$ & $54.79_{-2.08}^{+2.45}$ & $11.98_{-0.8}^{+0.88}$ & $3.3_{-0.81}^{+1.07}$ & $1.22_{-0.02}^{+0.03}$ & $3.24_{-0.21}^{+0.07}$ & $2.89_{-1.35}^{+1.6}$ & $13.65_{-0.24}^{+0.27}$ & $0.14_{-0.02}^{+0.02}$ & $0.29_{-0.01}^{+0.01}$ & $0.135_{-0.006}^{+0.006}$ & $1.22_{-0.008}^{+0.008}$ & $1.969_{-0.046}^{+0.046}$ & \\
6203980105 &60185.04 &   & &  & &  &  & & & &  & &  & &  &  &  & \\
& &   & &  & &  &  & & & &  & &  & &  &  &  & \\
\hline

P061433800405&60187.70 &  $0.21_{-0.01}^{+0.01}$ & $0.45_{-0.01}^{+0.01}$ & $7.35_{-0.41}^{+0.74}$ & $1.77_{-0.03}^{+0.01}$ & $64.28_{-3.17}^{+0.96}$ & $15.57_{-1.07}^{+1.08}$ & $3.44_{-0.92}^{+2.92}$ & $1.29_{-0.03}^{+0.03}$ & $3.29_{-0.3}^{+0.05}$ & $2.48_{-1.62}^{+0.65}$ & $14.23_{-0.54}^{+0.2}$ & $0.13_{-0.002}^{+0.005}$ & $0.28_{-0.01}^{+0.01}$ & $0.149_{-0.003}^{+0.003}$ & $1.251_{-0.009}^{+0.009}$ & $1.807_{-0.042}^{+0.042}$ & \\
6203980107 &60187.74&   & &  & &  &  & & & &  & &  & &  &  &  & \\
& &   & &  & &  &  & & & &  & &  & &  &  &  & \\
\hline

P061433800407&60187.97 &  $0.23_{-0.01}^{+0.01}$ & $0.46_{-0.01}^{+0.01}$ & $5.99_{-0.74}^{+0.39}$ & $1.9_{-0.03}^{+0.01}$ & $66.47_{-4.57}^{+0.78}$ & $27.37_{-2.03}^{+1.86}$ & $2.51_{-0.18}^{+0.15}$ & $1.06_{-0.05}^{+0.04}$ & $3.24_{-0.22}^{+0.14}$ & $10.0_{-5.28}^{+0.0}$ & $11.43_{-0.41}^{+0.33}$ & $0.14_{-0.005}^{+0.025}$ & $0.19_{-0.01}^{+0.01}$ & $0.146_{-0.003}^{+0.003}$ & $1.692_{-0.012}^{+0.012}$ & $1.329_{-0.063}^{+0.063}$ & \\
6203980108 &60188.01&   & &  & &  &  & & & &  & &  & &  &  &  & \\
& &   & &  & &  &  & & & &  & &  & &  &  &  & \\
\hline

P061433800501&60189.08 &  $0.23_{-0.01}^{+0.01}$ & $0.45_{-0.01}^{+0.01}$ & $7.36_{-0.64}^{+1.03}$ & $1.79_{-0.03}^{+0.01}$ & $63.19_{-2.02}^{+1.85}$ & $16.85_{-1.22}^{+0.79}$ & $2.3_{-0.0}^{+0.0}$ & $1.3_{-0.01}^{+0.03}$ & $3.31_{-0.15}^{+0.14}$ & $10.0_{-3.6}^{+0.0}$ & $13.79_{-0.29}^{+0.37}$ & $0.1_{-0.005}^{+0.01}$ & $0.26_{-0.007}^{+0.01}$ & $0.152_{-0.004}^{+0.003}$ & $1.283_{-0.012}^{+0.012}$ & $1.634_{-0.038}^{+0.038}$ & \\
6203980109  &60189.04&   & &  & &  &  & & & &  & &  & &  &  &  & \\
& &   & &  & &  &  & & & &  & &  & &  &  &  & \\
\hline

P061433800508&60190.02 &  $0.25_{-0.01}^{+0.01}$ & $0.45_{-0.01}^{+0.01}$ & $6.92_{-1.04}^{+1.04}$ & $1.92_{-0.02}^{+0.01}$ & $65.87_{-1.06}^{+1.36}$ & $27.18_{-2.21}^{+0.84}$ & $2.5_{-0.0}^{+0.0}$ & $1.17_{-0.03}^{+0.04}$ & $3.29_{-0.2}^{+0.07}$ & $10.0_{-2.42}^{+0.0}$ & $12.32_{-0.22}^{+0.32}$ & $0.15_{-0.005}^{+0.01}$ & $0.19_{-0.01}^{+0.01}$ & $0.139_{-0.003}^{+0.003}$ & $1.59_{-0.015}^{+0.015}$ & $1.285_{-0.03}^{+0.03}$ & \\
6750010201  &60190.01&   & &  & &  &  & & & &  & &  & &  &  &  & \\
& &   & &  & &  &  & & & &  & &  & &  &  &  & \\
\hline

P061433800512&60190.55 &  $0.22_{-0.02}^{+0.01}$ & $0.46_{-0.01}^{+0.01}$ & $6.94_{-1.03}^{+0.44}$ & $1.74_{-0.02}^{+0.03}$ & $67.94_{-3.06}^{+3.2}$ & $11.46_{-0.91}^{+0.94}$ & $3.44_{-0.65}^{+2.51}$ & $1.42_{-0.01}^{+0.02}$ & $3.29_{-0.14}^{+0.2}$ & $4.79_{-2.71}^{+0.7}$ & $15.95_{-0.35}^{+0.33}$ & $0.07_{-0.004}^{+0.01}$ & $0.31_{-0.01}^{+0.005}$ & $0.151_{-0.004}^{+0.003}$ & $1.007_{-0.009}^{+0.009}$ & $1.856_{-0.043}^{+0.043}$ & \\
6203980110  &60190.52&   & &  & &  &  & & & &  & &  & &  &  &  & \\
& &   & &  & &  &  & & & &  & &  & &  &  &  & \\
\hline

P061433800603&60191.34 &  $0.2_{-0.01}^{+0.01}$ & $0.45_{-0.01}^{+0.01}$ & $6.46_{-0.90}^{+0.91}$ & $1.75_{-0.02}^{+0.02}$ & $59.85_{-2.72}^{+3.98}$ & $15.79_{-0.76}^{+1.33}$ & $3.5_{-1.08}^{+0.59}$ & $1.25_{-0.03}^{+0.03}$ & $3.59_{-0.17}^{+0.2}$ & $6.23_{-0.69}^{+2.58}$ & $13.7_{-0.32}^{+0.45}$ & $0.12_{-0.01}^{+0.01}$ & $0.23_{-0.01}^{+0.01}$ & $0.135_{-0.003}^{+0.003}$ & $1.314_{-0.012}^{+0.012}$ & $1.488_{-0.035}^{+0.035}$ & \\
6703010103 &60191.35&   & &  & &  &  & & & &  & &  & &  &  &  & \\
& &   & &  & &  &  & & & &  & &  & &  &  &  & \\
\hline

P061433800608&60192.00 &  $0.21_{-0.01}^{+0.02}$ & $0.46_{-0.01}^{+0.02}$ & $5.73_{-1.06}^{+0.92}$ & $1.74_{-0.02}^{+0.03}$ & $79.81_{-5.1}^{+8.46}$ & $8.71_{-0.32}^{+0.87}$ & $3.1_{-1.35}^{+1.46}$ & $1.51_{-0.03}^{+0.01}$ & $3.04_{-0.15}^{+0.04}$ & $1.04_{-0.14}^{+1.16}$ & $19.11_{-0.71}^{+0.43}$ & $0.15_{-0.01}^{+0.02}$ & $0.31_{-0.01}^{+0.01}$ & $0.13_{-0.003}^{+0.003}$ & $0.801_{-0.007}^{+0.007}$ & $1.918_{-0.045}^{+0.045}$ & \\
6203980112  &60192.06&   & &  & &  &  & & & &  & &  & &  &  &  & \\
& &   & &  & &  &  & & & &  & &  & &  &  &  & \\
\hline

P061433800616&60193.06 &  $0.22_{-0.01}^{+0.02}$ & $0.47_{-0.01}^{+0.03}$ & $5.65_{-0.16}^{+0.93}$ & $1.8_{-0.02}^{+0.03}$ & $70.7_{-4.21}^{+4.91}$ & $13.8_{-1.11}^{+1.3}$ & $3.0_{-0.0}^{+0.0}$ & $1.43_{-0.02}^{+0.03}$ & $3.42_{-0.26}^{+0.22}$ & $10.0_{-3.12}^{+0.0}$ & $15.83_{-0.51}^{+0.65}$ & $0.09_{-0.005}^{+0.005}$ & $0.26_{-0.02}^{+0.02}$ & $0.145_{-0.003}^{+0.003}$ & $1.059_{-0.007}^{+0.007}$ & $1.548_{-0.036}^{+0.036}$ & \\
6750010301 &60193.03&   & &  & &  &  & & & &  & &  & &  &  &  & \\
& &   & &  & &  &  & & & &  & &  & &  &  &  & \\
\hline

P061433800618 &60193.32 &  $0.21_{-0.01}^{+0.01}$ & $0.47_{-0.01}^{+0.01}$ & $5.35_{-0.68}^{+0.79}$ & $1.81_{-0.01}^{+0.04}$ & $76.47_{-2.21}^{+9.3}$ & $12.84_{-1.34}^{+1.02}$ & $2.97_{-0.99}^{+1.47}$ & $1.43_{-0.03}^{+0.03}$ & $3.3_{-0.17}^{+0.34}$ & $4.05_{-1.91}^{+1.01}$ & $16.45_{-0.54}^{+0.74}$ & $0.1_{-0.01}^{+0.025}$ & $0.26_{-0.02}^{+0.01}$ & $0.142_{-0.003}^{+0.003}$ & $1.007_{-0.007}^{+0.007}$ & $1.614_{-0.038}^{+0.038}$ & \\
6203980112 &60193.42&   & &  & &  &  & & & &  & &  & &  &  &  & \\
& &   & &  & &  &  & & & &  & &  & &  &  &  & \\
\hline

P061433800801&60194.04 &  $0.22_{-0.01}^{+0.01}$ & $0.46_{-0.01}^{+0.01}$ & $5.52_{-0.70}^{+0.78}$ & $1.84_{-0.02}^{+0.02}$ & $70.14_{-3.1}^{+3.46}$ & $17.0_{-1.17}^{+0.82}$ & $3.0_{-0.0}^{+0.0}$ & $1.39_{-0.03}^{+0.02}$ & $3.31_{-0.12}^{+0.2}$ & $3.1_{-0.72}^{+0.86}$ & $16.1_{-0.56}^{+0.41}$ & $0.15_{-0.01}^{+0.02}$ & $0.22_{-0.01}^{+0.01}$ & $0.131_{-0.003}^{+0.003}$ & $1.196_{-0.008}^{+0.008}$ & $1.409_{-0.033}^{+0.033}$ & \\
6203980114 &60194.00&   & &  & &  &  & & & &  & &  & &  &  &  & \\
& &   & &  & &  &  & & & &  & &  & &  &  &  & \\
\hline

P061433800905 &60195.64 &  $0.21_{-0.01}^{+0.02}$ & $0.46_{-0.01}^{+0.01}$ & $5.47_{-0.54}^{+0.68}$ & $1.77_{-0.02}^{+0.03}$ & $63.87_{-2.8}^{+3.78}$ & $14.3_{-1.26}^{+1.05}$ & $2.83_{-0.96}^{+0.97}$ & $1.36_{-0.04}^{+0.04}$ & $3.32_{-0.31}^{+0.27}$ & $3.61_{-1.86}^{+1.82}$ & $15.18_{-0.75}^{+0.73}$ & $0.18_{-0.015}^{+0.005}$ & $0.2_{-0.01}^{+0.01}$ & $0.13_{-0.006}^{+0.006}$ & $1.158_{-0.013}^{+0.013}$ & $1.302_{-0.061}^{+0.061}$ & \\
6203980115 &60195.61&   & &  & &  &  & & & &  & &  & &  &  &  & \\
& &   & &  & &  &  & & & &  & &  & &  &  &  & \\
\hline

    \hline
        \label{T-R-H} &     

    \end{tabular}
}

\end{table*}

\begin{table*}[]
    \centering
		\caption{The results of the joint fit, corresponding to Figure \ref{rela-tail}.}
    \resizebox{\textwidth}{0.5\textwidth}{
		\begin{tabular}{ccccccccccccccccccc}
            
		  \hline
		   \hline
          Insight-HXMT & MJD&$N_{\rm H}$&$T_{\rm in}$&$N_{\rm disk}$ &PhoIndex & HighEcut&$N_{\rm cutoffpl}$&$R_{\rm in}$&$\Gamma$&logxi&Afe&Ecut&feflfrac&$N_{\rm relxill}$&$F_{\rm disk}$&$F_{\rm cutoffpl}$&$F_{\rm relxill}$
          \\NICER& &&& && &&&&&&&&&&&
     \\NuSTAR& &&& && &&&&&&&&&&&  
         \\ObsID &&$(10^{22} \rm cm^{-2})$&(keV)&($10^{4})$&&(keV)&&($R_{\rm g}$)&&&&(keV)&&&($10^{-8}$~erg~s$^{-1}$~cm$^{-2}$)&$(10^{-8}$~erg~s$^{-1}$~cm$^{-2}$)&$(10^{-8}$~erg~s$^{-1}$~cm$^{-2}$)
       \\ \hline

P061433800302&60185.40 & $0.16_{-0.01}^{+0.01}$ & $0.46_{-0.01}^{+0.01}$ & $6.37_{-0.56}^{+0.80}$ & $1.22_{-0.02}^{+0.02}$ & $12.96_{-0.31}^{+0.09}$ & $17.54_{-1.07}^{+0.72}$ & $2.47_{-0.17}^{+0.32}$ & $1.53_{-0.03}^{+0.03}$ & $3.16_{-0.09}^{+0.18}$ & $0.90_{-0.23}^{+0.11}$ & $51.27_{-1.18}^{+1.94}$ & $0.51_{-0.07}^{+0.10}$ & $0.18_{-0.01}^{+0.01}$ & $0.14_{-0.005}^{+0.004}$ & $1.85_{-0.009}^{+0.009}$ & $1.39_{-0.03}^{+0.04}$ & \\ 
6703010101&60185.55 &  
  & &  & &  &  & & & &  & &  & &  &  &  & \\
90902330002&60185.42 &   & &  & &  &  & & & &  & &  & &  &  &  & \\
\hline

P0614330800411&60188.50 &  $0.21_{-0.01}^{+0.02}$ & $0.47_{-0.01}^{+0.01}$ & $6.71_{-0.51}^{+0.98}$ & $1.34_{-0.01}^{+0.02}$  & $13.34_{-0.1}^{+0.02}$& $19.41_{-0.85}^{+0.73}$ & $2.27_{-0.31}^{+0.22}$ & $1.62_{-0.02}^{+0.03}$ & $3.14_{-0.06}^{+0.06}$ & $0.87_{-0.16}^{+0.10}$ & $57.14_{-2.79}^{+3.85}$ & $0.48_{-0.05}^{+0.05}$ & $0.18_{-0.01}^{+0.01}$ & $0.17_{-0.004}^{+0.004}$ & $1.64_{-0.008}^{+0.008}$ & $1.35_{-0.03}^{+0.03}$ & \\ 
6703010102&60185.58 &   & &  & &  &  & & & &  & &  & &  &  &  & \\
80902330002&60185.56 &   & &  & &  &  & & & &  & &  & &  &  &  & \\
\hline

P0614330800601&60191.06 &  $0.18_{-0.01}^{+0.01}$ & $0.48_{-0.01}^{+0.01}$ & $5.40_{-0.57}^{+0.54}$ & $1.37_{-0.01}^{+0.01}$ & $14.36_{-0.27}^{+0.23}$ & $20.39_{-0.40}^{+0.79}$ & $1.99_{-0.43}^{+0.21}$ & $1.51_{-0.04}^{+0.01}$ & $3.16_{-0.05}^{+0.04}$ & $1.29_{-0.19}^{+0.62}$ & $61.09_{-2.80}^{+0.74}$ & $0.64_{-0.08}^{+0.03}$ & $0.13_{-0.01}^{+0.01}$ & $0.15_{-0.004}^{+0.003}$ & $1.74_{-0.008}^{+0.008}$ & $1.10_{-0.03}^{+0.03}$ & \\ 
6203980111&60191.04 &   & &  & &  &  & & & &  & &  & &  &  &  & \\
80902330004&60191.10 &   & &  & &  &  & & & &  & &  & &  &  &  & \\
\hline

P0614330800807&60194.85 &  $0.18_{-0.01}^{+0.01}$ & $0.5_{-0.01}^{+0.01}$ & $4.41_{-0.46}^{+0.47}$ & $1.48_{-0.02}^{+0.01}$ & $13.62_{-0.90}^{+0.30}$ & $24.29_{-1.73}^{+1.33}$ & $1.61_{-0.12}^{+0.61}$ & $1.40_{-0.07}^{+0.03}$ & $3.07_{-0.07}^{+0.24}$ & $1.02_{-0.09}^{+0.55}$ & $60.09_{-0.7.42}^{+4.69}$ & $0.94_{-0.11}^{+0.19}$ & $0.19_{-0.01}^{+0.01}$ & $0.15_{-0.004}^{+0.004}$ & $1.63_{-0.008}^{+0.007}$ & $0.88_{-0.04}^{+0.04}$ & \\ 
6750010501&60194.79 &   & &  & &  &  & & & &  & &  & &  &  &  & \\
80902330006&60194.78 &   & &  & &  &  & & & &  & &  & &  &  &  & \\
\hline

P0614330800901&60195.09 &  $0.19_{-0.01}^{+0.01}$ & $0.48_{-0.01}^{+0.01}$ & $4.84_{-0.19}^{+0.19}$ & $1.43_{-0.04}^{+0.02}$ & $15.69_{-0.38}^{+0.43}$ & $17.47_{-1.57}^{+0.98}$ & $2.51_{-0.38}^{+0.30}$ & $1.69_{-0.04}^{+0.04}$ & $3.41_{-0.11}^{+0.16}$ & $0.90_{-0.31}^{+0.20}$ & $77.79_{-6.01}^{+5.64}$ & $0.68_{-0.15}^{+0.10}$ & $0.13_{-0.01}^{+0.01}$ & $0.18_{-0.004}^{+0.004}$ & $1.40_{-0.006}^{+0.006}$ & $0.88_{-0.04}^{+0.04}$ & \\ 
6750010502&60195.03 &   & &  & &  &  & & & &  & &  & &  &  &  & \\
80902330008&60195.05 &   & &  & &  &  & & & &  & &  & &  &  &  & \\
\hline

P061433800101&60181.34 &  $0.24_{-0.06}^{+0.06}$ & $0.28_{-0.01}^{+0.02}$ & $16.94_{-7.12}^{+8.66}$ & $0.0_{-0.0}^{+0.0}$ & $0.0_{-0.0}^{+0.0}$ & $0.0_{-0.0}^{+0.0}$ & $2.21_{-0.44}^{+0.32}$ & $1.48_{-0.02}^{+0.02}$ & $2.85_{-0.06}^{+0.07}$ & $0.57_{-0.03}^{+0.06}$ & $46.15_{-1.33}^{+0.82}$ & $0.47_{-0.01}^{+0.015}$ & $0.23_{-0.005}^{+0.004}$ & $0.024_{-0.005}^{+0.004}$ & $0_{-0.0}^{+0.0}$ & $1.699_{-0.008}^{+0.008}$ & \\ 
6203980101&60181.03 &   & &  & &  &  & & & &  & &  & &  &  &  & \\
& &   & &  & &  &  & & & &  & &  & &  &  &  & \\
\hline

P061433800110&60182.56&  $0.13_{-0.01}^{+0.02}$ & $0.4_{-0.01}^{+0.01}$ & $5.48_{-0.87}^{+0.96}$ & $1.16_{-0.01}^{+0.02}$ & $15.29_{-0.42}^{+0.24}$ & $11.48_{-0.4}^{+0.67}$ & $3.27_{-1.42}^{+0.51}$ & $1.49_{-0.06}^{+0.06}$ & $3.19_{-0.19}^{+0.08}$ & $1.25_{-0.2}^{+0.79}$ & $61.81_{-5.52}^{+3.98}$ & $0.52_{-0.04}^{+0.035}$ & $0.12_{-0.01}^{+0.01}$ & $0.053_{-0.004}^{+0.004}$ & $1.609_{-0.007}^{+-0.007}$ & $0.964_{-0.018}^{+0.018}$ & \\ 
6203980102&60182.59 &   & &  & &  &  & & & &  & &  & &  &  &  & \\
& &   & &  & &  &  & & & &  & &  & &  &  &  & \\
\hline

P061433800203&60183.35 &  $0.19_{-0.02}^{+0.02}$ & $0.35_{-0.01}^{+0.01}$ & $13.72_{-2.41}^{+2.53}$ & $1.26_{-0.02}^{+0.02}$ & $14.94_{-0.47}^{+0.61}$ & $13.23_{-0.42}^{+0.66}$ & $2.29_{-0.64}^{+0.34}$ & $1.52_{-0.04}^{+0.03}$ & $2.95_{-0.1}^{+0.07}$ & $0.84_{-0.12}^{+0.16}$ & $58.85_{-4.48}^{+4.55}$ & $0.59_{-0.055}^{+0.07}$ & $0.1_{-0.007}^{+0.009}$ & $0.068_{-0.007}^{+0.005}$ & $1.451_{-0.007}^{+-0.007}$ & $0.821_{-0.011}^{+0.011}$ & \\ 
6203980103&60183.36 &   & &  & &  &  & & & &  & &  & &  &  &  & \\
& &   & &  & &  &  & & & &  & &  & &  &  &  & \\
\hline

P061433800212&60184.54 &  $0.17_{-0.01}^{+0.01}$ & $0.42_{-0.01}^{+0.01}$ & $8.98_{-0.70}^{+0.98}$ & $1.25_{-0.02}^{+0.02}$ & $12.81_{-0.35}^{+0.68}$ & $22.13_{-1.04}^{+0.73}$ & $2.71_{-0.8}^{+0.68}$ & $1.43_{-0.03}^{+0.05}$ & $2.91_{-0.13}^{+0.16}$ & $1.24_{-0.35}^{+0.67}$ & $51.59_{-1.55}^{+3.51}$ & $0.62_{-0.045}^{+0.04}$ & $0.13_{-0.01}^{+0.01}$ & $0.133_{-0.006}^{+0.006}$ & $2.151_{-0.01}^{+0.01}$ & $1.069_{-0.017}^{+0.017}$ & \\ 
6203980104&60184.65 &   & &  & &  &  & & & &  & &  & &  &  &  & \\
& &   & &  & &  &  & & & &  & &  & &  &  &  & \\
\hline

P061433800301&60185.30 &  $0.19_{-0.01}^{+0.02}$ & $0.42_{-0.01}^{+0.01}$ & $9.57_{-1.34}^{+1.60}$ & $1.22_{-0.01}^{+0.02}$ & $12.97_{-0.15}^{+0.43}$ & $17.34_{-0.61}^{+0.81}$ & $3.28_{-0.65}^{+1.11}$ & $1.64_{-0.01}^{+0.03}$ & $3.26_{-0.16}^{+0.12}$ & $2.23_{-0.53}^{+0.42}$ & $54.78_{-0.73}^{+2.46}$ & $0.29_{-0.02}^{+0.02}$ & $0.2_{-0.01}^{+0.01}$ & $0.134_{-0.006}^{+0.006}$ & $1.817_{-0.008}^{+0.008}$ & $1.368_{-0.016}^{+0.016}$ & \\ 
6203980105 &60185.04 &   & &  & &  &  & & & &  & &  & &  &  &  & \\
& &   & &  & &  &  & & & &  & &  & &  &  &  & \\
\hline

P061433800405&60187.70&  $0.2_{-0.01}^{+0.02}$ & $0.46_{-0.01}^{+0.01}$ & $7.45_{-0.99}^{+1.19}$ & $1.41_{-0.02}^{+0.02}$ & $14.38_{-0.56}^{+0.16}$ & $26.74_{-2.21}^{+0.62}$ & $3.3_{-0.0}^{+0.0}$ & $1.41_{-0.02}^{+0.06}$ & $3.57_{-0.13}^{+0.25}$ & $2.36_{-1.06}^{+1.38}$ & $57.85_{-1.32}^{+2.57}$ & $0.7_{-0.035}^{+0.115}$ & $0.1_{-0.01}^{+0.01}$ & $0.16_{-0.004}^{+0.004}$ & $2.096_{-0.01}^{+0.01}$ & $0.931_{-0.011}^{+0.011}$ & \\ 
6203980107 &60187.74&   & &  & &  &  & & & &  & &  & &  &  &  & \\
& &   & &  & &  &  & & & &  & &  & &  &  &  & \\
\hline

P061433800407&60187.97&  $0.21_{-0.01}^{+0.01}$ & $0.47_{-0.01}^{+0.01}$ & $6.33_{-0.70}^{+0.70}$ & $1.34_{-0.02}^{+0.03}$ & $13.78_{-0.43}^{+0.31}$ & $21.39_{-0.95}^{+1.29}$ & $2.7_{-0.51}^{+1.66}$ & $1.75_{-0.04}^{+0.04}$ & $3.17_{-0.23}^{+0.15}$ & $4.97_{-2.08}^{+0.62}$ & $77.95_{-5.75}^{+6.77}$ & $0.25_{-0.03}^{+0.045}$ & $0.18_{-0.02}^{+0.02}$ & $0.162_{-0.004}^{+0.004}$ & $1.882_{-0.009}^{+0.009}$ & $1.075_{-0.025}^{+0.025}$ & \\ 
6203980108 &60188.01&   & &  & &  &  & & & &  & &  & &  &  &  & \\
& &   & &  & &  &  & & & &  & &  & &  &  &  & \\
\hline

P061433800501&60189.08  &  $0.2_{-0.007}^{+0.01}$ & $0.47_{-0.01}^{+0.01}$ & $6.81_{-0.33}^{+0.54}$ & $1.42_{-0.01}^{+0.02}$ & $14.18_{-0.34}^{+0.29}$ & $26.28_{-0.87}^{+0.95}$ & $3.01_{-0.64}^{+0.8}$ & $1.41_{-0.04}^{+0.03}$ & $3.59_{-0.07}^{+0.06}$ & $6.63_{-1.95}^{+2.64}$ & $53.99_{-2.23}^{+1.46}$ & $0.35_{-0.025}^{+0.015}$ & $0.11_{-0.01}^{+0.01}$ & $0.166_{-0.004}^{+0.004}$ & $2.029_{-0.009}^{+0.009}$ & $0.843_{-0.008}^{+0.008}$ & \\ 
6203980109  &60189.04&   & &  & &  &  & & & &  & &  & &  &  &  & \\
& &   & &  & &  &  & & & &  & &  & &  &  &  & \\
\hline

P061433800508&60190.02 &  $0.2_{-0.01}^{+0.02}$ & $0.47_{-0.02}^{+0.01}$ & $6.30_{-0.69}^{+1.10}$ & $1.45_{-0.02}^{+0.03}$ & $14.44_{-0.55}^{+0.44}$ & $27.92_{-0.6}^{+1.47}$ & $2.5_{-0.0}^{+0.0}$ & $1.4_{-0.04}^{+0.05}$ & $3.54_{-0.23}^{+0.03}$ & $9.25_{-4.35}^{+0.25}$ & $56.99_{-3.01}^{+2.56}$ & $0.43_{-0.04}^{+0.035}$ & $0.09_{-0.01}^{+0.01}$ & $0.166_{-0.004}^{+0.004}$ & $2.04_{-0.009}^{+0.009}$ & $0.747_{-0.012}^{+0.012}$ & \\ 
6750010201  &60190.01&   & &  & &  &  & & & &  & &  & &  &  &  & \\
& &   & &  & &  &  & & & &  & &  & &  &  &  & \\
\hline

P061433800512&60190.55  &  $0.2_{-0.02}^{+0.01}$ & $0.47_{-0.01}^{+0.01}$ & $6.44_{-0.64}^{+0.49}$ & $1.43_{-0.02}^{+0.02}$ & $14.95_{-0.48}^{+0.43}$ & $23.61_{-1.22}^{+1.08}$ & $3.51_{-0.72}^{+0.6}$ & $1.61_{-0.03}^{+0.04}$ & $3.29_{-0.21}^{+0.11}$ & $2.88_{-0.56}^{+0.81}$ & $60.81_{-2.96}^{+4.4}$ & $0.28_{-0.01}^{+0.02}$ & $0.15_{-0.01}^{+0.01}$ & $0.157_{-0.004}^{+0.004}$ & $1.846_{-0.009}^{+0.008}$ & $0.993_{-0.011}^{+0.011}$ & \\ 
6203980110  &60190.52&   & &  & &  &  & & & &  & &  & &  &  &  & \\
& &   & &  & &  &  & & & &  & &  & &  &  &  & \\
\hline

P061433800603&60191.34&  $0.19_{-0.01}^{+0.01}$ & $0.46_{-0.01}^{+0.01}$ & $6.61_{-0.93}^{+0.87}$ & $1.35_{-0.01}^{+0.02}$ & $13.98_{-0.32}^{+0.67}$ & $19.23_{-0.82}^{+1.22}$ & $3.84_{-1.04}^{+2.42}$ & $1.6_{-0.02}^{+0.03}$ & $3.24_{-0.16}^{+0.12}$ & $3.06_{-0.92}^{+0.68}$ & $57.95_{-2.75}^{+5.31}$ & $0.29_{-0.025}^{+0.02}$ & $0.16_{-0.01}^{+0.01}$ & $0.142_{-0.003}^{+0.003}$ & $1.673_{-0.008}^{+0.008}$ & $1.109_{-0.015}^{+0.015}$ & \\ 
6703010103 &60191.35&   & &  & &  &  & & & &  & &  & &  &  &  & \\
& &   & &  & &  &  & & & &  & &  & &  &  &  & \\
\hline

P061433800608&60192.00 &  $0.21_{-0.01}^{+0.01}$ & $0.46_{-0.01}^{+0.01}$ & $5.80_{-0.61}^{+0.43}$ & $1.44_{-0.02}^{+0.02}$ & $16.05_{-0.38}^{+0.59}$ & $19.21_{-0.81}^{+1.0}$ & $3.2_{-0.77}^{+0.78}$ & $1.77_{-0.03}^{+0.03}$ & $3.12_{-0.15}^{+0.13}$ & $0.89_{-0.24}^{+0.19}$ & $72.05_{-5.64}^{+5.11}$ & $0.41_{-0.04}^{+0.045}$ & $0.18_{-0.08}^{+0.09}$ & $0.13_{-0.003}^{+0.003}$ & $1.551_{-0.007}^{+0.007}$ & $1.167_{-0.016}^{+0.016}$ & \\ 
6203980112  &60192.06&   & &  & &  &  & & & &  & &  & &  &  &  & \\
& &   & &  & &  &  & & & &  & &  & &  &  &  & \\
\hline

P061433800616&60193.06 &  $0.2_{-0.02}^{+0.01}$ & $0.48_{-0.01}^{+0.01}$ & $5.66_{-0.83}^{+0.58}$ & $1.51_{-0.02}^{+0.02}$ & $15.19_{-0.79}^{+0.33}$ & $26.4_{-1.07}^{+1.1}$ & $3.0_{-0.0}^{+0.0}$ & $1.4_{-0.05}^{+0.04}$ & $3.61_{-0.12}^{+0.07}$ & $6.44_{-1.99}^{+0.08}$ & $54.79_{-4.04}^{+0.94}$ & $0.41_{-0.025}^{+0.025}$ & $0.1_{-0.01}^{+0.01}$ & $0.154_{-0.007}^{+0.003}$ & $1.803_{-0.008}^{+0.008}$ & $0.769_{-0.009}^{+0.009}$ & \\ 
6750010301 &60193.03&   & &  & &  &  & & & &  & &  & &  &  &  & \\
& &   & &  & &  &  & & & &  & &  & &  &  &  & \\
\hline

P061433800618 &60193.32  &  $0.2_{-0.01}^{+0.01}$ & $0.47_{-0.02}^{+0.01}$ & $5.61_{-0.68}^{+0.93}$ & $1.49_{-0.01}^{+0.03}$ & $15.64_{-0.44}^{+0.54}$ & $24.28_{-0.83}^{+1.07}$ & $3.0_{-0.0}^{+0.0}$ & $1.57_{-0.04}^{+0.04}$ & $3.19_{-0.13}^{+0.15}$ & $2.26_{-0.82}^{+0.54}$ & $65.13_{-4.4}^{+5.81}$ & $0.42_{-0.025}^{+0.035}$ & $0.12_{-0.01}^{+0.01}$ & $0.147_{-0.003}^{+0.003}$ & $1.737_{-0.008}^{+0.008}$ & $0.864_{-0.012}^{+0.012}$ & \\ 
6203980112 &60193.42&   & &  & &  &  & & & &  & &  & &  &  &  & \\
& &   & &  & &  &  & & & &  & &  & &  &  &  & \\
\hline

P061433800801&60194.04&  $0.21_{-0.01}^{+0.01}$ & $0.47_{-0.01}^{+0.01}$ & $5.52_{-0.81}^{+0.69}$ & $1.44_{-0.02}^{+0.01}$ & $15.58_{-0.28}^{+0.19}$ & $17.84_{-0.93}^{+0.66}$ & $3.0_{-0.0}^{+0.0}$ & $1.76_{-0.03}^{+0.02}$ & $3.3_{-0.12}^{+0.16}$ & $2.12_{-0.53}^{+0.53}$ & $67.58_{-2.81}^{+1.87}$ & $0.28_{-0.02}^{+0.015}$ & $0.19_{-0.01}^{+0.01}$ & $0.134_{-0.003}^{+0.003}$ & $1.418_{-0.007}^{+0.007}$ & $1.173_{-0.014}^{+0.014}$ & \\ 
6203980114 &60194.00&   & &  & &  &  & & & &  & &  & &  &  &  & \\
& &   & &  & &  &  & & & &  & &  & &  &  &  & \\
\hline

P061433800905 &60195.64  &  $0.19_{-0.01}^{+0.01}$ & $0.48_{-0.01}^{+0.01}$ & $5.00_{-0.05}^{+0.07}$ & $1.36_{-0.02}^{+0.02}$ & $14.03_{-0.2}^{+0.35}$ & $14.45_{-0.91}^{+1.39}$ & $3.07_{-0.25}^{+0.98}$ & $1.63_{-0.03}^{+0.03}$ & $4.11_{-0.12}^{+0.03}$ & $10.0_{-4.52}^{+0.0}$ & $59.08_{-0.73}^{+1.77}$ & $0.36_{-0.04}^{+0.005}$ & $0.16_{-0.008}^{+0.008}$ & $0.139_{-0.003}^{+0.003}$ & $1.222_{-0.008}^{+0.008}$ & $1.2_{-0.022}^{+0.022}$ & \\ 
6203980115 &60195.61&   & &  & &  &  & & & &  & &  & &  &  &  & \\
& &   & &  & &  &  & & & &  & &  & &  &  &  & \\
\hline

    \hline
        \label{T-R-T} &     

    \end{tabular}
}
\end{table*}


\bibliography{sample631}{}
\bibliographystyle{aasjournal}



\end{document}